\documentclass[aps,prc,twocolumn,superscriptaddress,showpacs]{revtex4}
\usepackage{graphicx}
\usepackage{amsmath}
\usepackage[dvipsnames]{xcolor}

\newcommand{\beq}{\begin{equation}}
\newcommand{\eeq}{\end{equation}}

\begin{document}

\title{Gamma strength function and level density of $^{208}$Pb from forward-angle proton scattering at 295 MeV}

\newcommand{\RCNP}{Research Center for Nuclear Physics, Osaka University, Ibaraki, Osaka 567-0047, Japan}
\newcommand{\TUDarmstadt}{Institut f\"ur Kernphysik, Technische Universit\"{a}t Darmstadt, D-64289 Darmstadt, Germany}

\author{S.~Bassauer}\email{sbassauer@ikp.tu-darmstadt.de}\affiliation{\TUDarmstadt}
\author{P.~von~Neumann-Cosel}\email{vnc@ikp.tu-darmstadt.de}\affiliation{\TUDarmstadt}
\author{A.~Tamii}\affiliation{\RCNP}

\date{\today}

\begin{abstract}
\begin{description}

\item[Background:] 
Gamma strength functions (GSFs) and level densities (LDs) are essential ingredients of statistical nuclear reaction theory with many applications in astrophysics, reactor design, and waste transmutation. 
\item[Purpose:]
The aim of the present work is a test of systematic parametrizations of the GSF recommended by the RIPL-3 data base for the case of $^{208}$Pb.
The upward GSF and LD in $^{208}$Pb are compared to $\gamma$ decay data from an Oslo-type experiment to examine the validity of the Brink-Axel (BA) hypothesis.    
\item[Methods:] 
The E1 and M1 parts of the total GSF are determined from high-resolution forward angle inelastic proton scattering data taken at 295 MeV at RCNP, Osaka, Japan.  
The total LD in $^{208}$Pb is derived from the $1^-$ LD extracted with a fluctuation analysis in the energy region of the isovector giant dipole resonance.   
\item[Results:] 
The E1 GSF is compared to parametrizations recommended by the RIPL-3 data base showing systematic deficiencies of all models in the energy region around neutron threshold.
The new data for the poorly known spinflip M1 resonance call for a substantial revision of the model suggested in RIPL-3.
The total GSF derived from the present data is larger in the PDR energy region than the Oslo data but the strong fluctuations due to the low LD resulting from the double shell closure of $^{208}$Pb prevent a conclusion on a possible violation of the BA hypothesis.
Using the parameters suggested by RIPL-3 for a description of the LD in $^{208}$Pb with the back-shifted Fermi gas  model, remarkable agreement between the two experiments spanning a wide excitation energy range is obtained.
\item[Conclusions:]
Systematic parametrizations of the E1 and M1 GSF parts need to be reconsidered at low excitation energies.
The good agreement of the LD provides an independent confirmation of the approach underlying the decomposition of GSF and LD in Oslo-type experiments.
\end{description}
\end{abstract}


\maketitle

\section{Introduction}

Gamma strength functions describe the average $\gamma$ decay behavior of a nucleus. 
They serve as input for applications of statistical nuclear theory in astrophysics \cite{arn07}, reactor design \cite{cha11}, and waste transmutation \cite{sal11}.
Although all electromagnetic multipoles contribute, the GSFs are usually dominated by the E1 component with smaller contributions from M1 strength.
Above particle threshold it is governed by the isovector giant dipole resonance (IVGDR) but for astrophysical processes the energy region around particle thresholds and at even lower excitation energies \cite{rau08} is more important.  
There, the situation is more complex:
In nuclei with neutron excess one observes the formation of the pygmy dipole resonance (PDR) \cite{sav13} but the low-energy tail of the IVGDR can also contribute.
Furthermore, the spinflip M1 resonance overlaps with the energy region of the PDR \cite{hey10}. 
The impact of the low-energy GSFs on astrophysical reaction rates and the resulting abundances in the $r$-process have been discussed e.g.\ in Refs. \cite{gor04,lit09,dao12}.

Many applications imply an environment of finite temperature, notably in stellar scenarios \cite{wie12}, and thus reactions on initially excited states become relevant.
Their contributions to the reaction rates are usually estimated applying the generalized Brink-Axel (BA) hypothesis \cite{bri55,axe62} which states that the GSF is independent of the properties of the initial and final states.
The validity of the BA hypothesis is also implicity assumed in the derivation of the GSFs from many experimental data based on ground state photoexcitation.  
Although historically formulated for the IVGDR, where it seems to hold approximately for not too high temperatures \cite{bbb98}, the BA hypothesis is nowadays commonly used to calculate the low-energy E1 and M1 strength functions. 
This is questioned by a recent shell-model analysis \cite{joh15} where it was demonstrated that the strength functions of collective modes built on excited states do show an energy dependence and this is expected from spectral distribution theory.
However, the numerical results for E1 strength functions showed an aproximate constancy consistent with the BA hypothesis. 

Recent work utilizing compound nucleus $\gamma$ decay with the so-called Oslo method \cite{sch00} has demonstrated independence of the GSF from excitation energies and spins of initial and final states in accordance with the BA hypothesis once the level densities are sufficiently high to suppress large intensity fluctuations \cite{gut16}.
However, there are a number of experimental results which seem to violate the BA hypothesis in the low-energy region.
For example, the GSFs in heavy deformed nuclei at excitation energies of $2 -3$ MeV are dominated by the orbital M1 scissors mode \cite{boh84} and large differences of $B$(M1) strengths are observed between $\gamma$ emission \cite{krt04,gut12} and absorption \cite{end05} experiments.      
For the low-energy E1 strength the question is far from clear when comparing results from the Oslo method with photoabsorption data.
Below particle thresholds much of the information on GSFs stems from nuclear resonance fluorescence (NRF) experiments.
A problem of the NRF method are experimentally unobserved branching ratios to excited states which have been neglected in many cases \cite{sav13}.
Recent studies of the $\gamma$ decay after photoabsoprtion indicate that these may be sizable \cite{rom15,loh16}. 

Here we present results for $^{208}$Pb from a new method for the measurement of E1 and M1 strength distributions in nuclei (and thus the GSF) from about 5 to 25 MeV based on relativistic Coulomb excitation in inelastic polarized proton scattering at energies of a few hundred MeV and scattering angles close to $0^\circ$ \cite{tam11,pol12,kru15,has15}.
The E1 strength distribution from Coulomb excitation permits to determine the dipole polarizability which provides important constraints on the neutron skin of nuclei and the poorly known parameters of the symmetry energy \cite{tam14}.
It also allows extraction of the M1 part of the GSF \cite{bir16} due to spinflip excitations which energetically overlaps with the PDR strength.
The high-resolution data also provide information on level densities -- another essential ingredient of statistical model cross section calculations -- from an analysis of the fine structure of the IVGDR \cite{pol14}.

The purpose of the paper is twofold. 
On one hand, recommended parametrizations of the GSF and LD summarized in the RIPL-3 data base \cite{cap09} are evaluated for the case of $^{208}$Pb.
In particular, we provide new data for the poorly known M1 part of the GSF.
On the other hand, the comparison with the Oslo experiment provides a test of the BA hypothesis.
Moreover, since GSF and LD are independently determined, the decomposition of both quantities in the Oslo method, which measures the product of GSF and LD \cite{sch00}, can be verified.      

\section{Gamma strength function of $^{208}\text{Pb}$}
\label{sec:gsf}

In the experiments discussed here, the GSF for an electric or magnetic transition $X\in{\left\lbrace E,M\right\rbrace }$ with multipolarity $\lambda$ is related to the photoabsoprtion cross section $\left\langle\sigma_{abs}^{X\lambda}\right\rangle$
\begin{equation}
	\label{eqn:gsf}
	f^{X\lambda}(E, J) = \frac{2J_0+1}{(\pi\hbar c)^2 (2J+1)}
	\frac{\left\langle\sigma_{abs}^{X\lambda}\right\rangle}{E^{2\lambda-1}} ,
\end{equation}
where  $E$ denotes the $\gamma$ energy and $J, J_0$ the spins of excited and ground state, respectively~\cite{cap09}.
The brackets $\langle \rangle$ indicate averaging over an energy interval.
In practise, only E1, M1 and E2 provide sizable contributions to the total GSF. 
In the following, we discuss the derivation of the GSF for these components from the experimental data and compare to parametrizations recommended in \mbox{RIPL-3}.

\subsection{E1 contribution}
\label{subsec:e1}

\begin{figure}
	\centering
	\includegraphics[width=8.6cm]{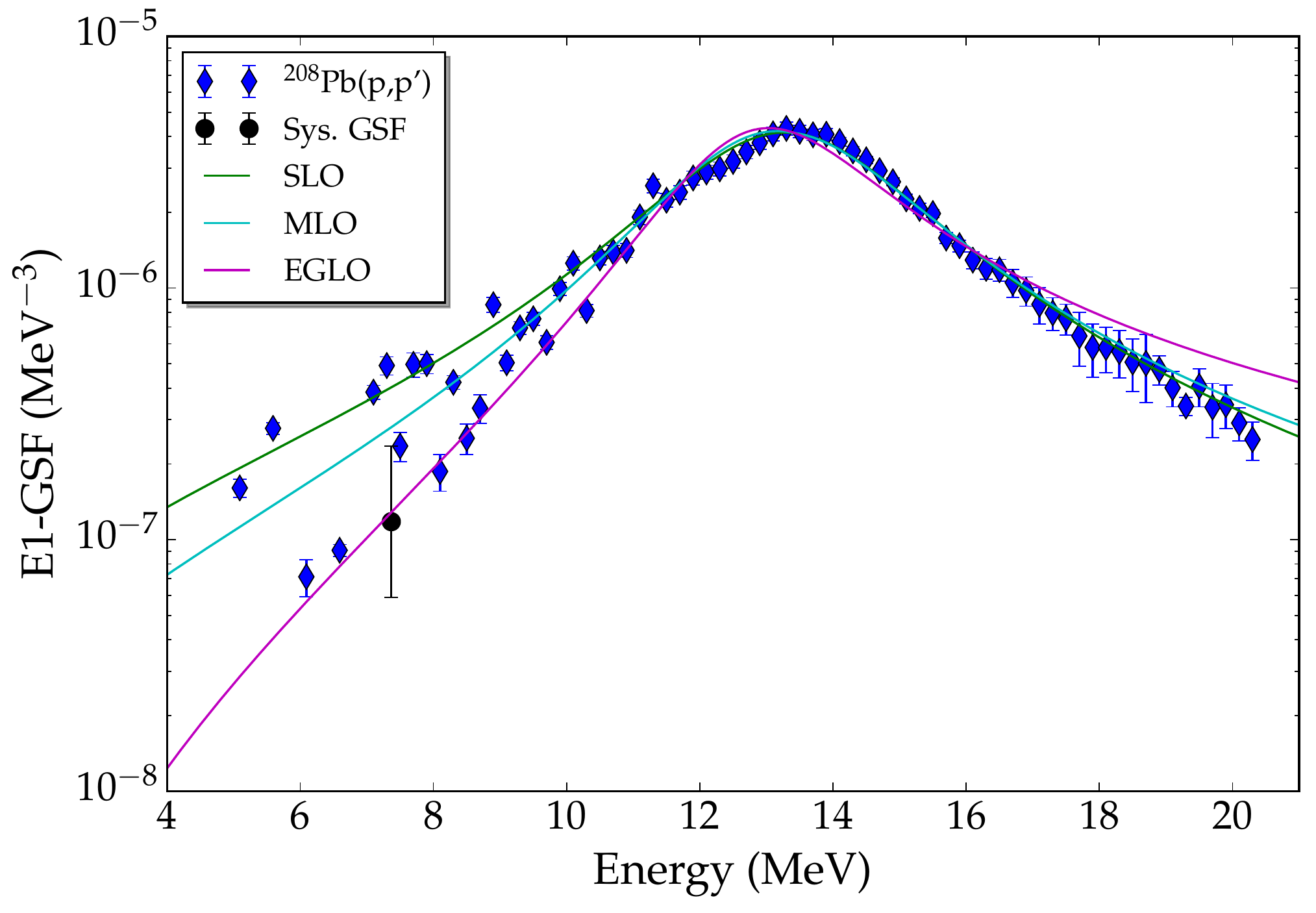}
	\caption{(Color online) E1 GSF of $^{208}$Pb deduced from the $(p,p^\prime)$ data \cite{tam11,pol12} (blue diamonds) in comparison with the SLO (green line), MLO (cyan line), and EGLO (magenta line) models explained in the text.
The black circle shows the prediction from experimental systematics at the neutron separation threshold \cite{cap09}.}
	\label{fig:e1}
\end{figure}

The E1 contribution of the GSF in $^{208}$Pb was determined using polarized inelastic proton scattering data obtained at the Research Center for Nuclear Physics (RCNP) at Osaka, Japan with a beam energy of 295~MeV in an excitation energy region from 5 to 23~MeV \cite{tam11, pol12}. 
In Ref.~\cite{pol12}, the $B$(E1) strength was extracted by means of the multipole decomposition analysis (MDA) in the energy region from 4.8 to 9~MeV. 
The B(E1) strength provided in Table I was used to determine the E1 part of the GSF. 
In the IVGDR region, the E1 GSF was determined from photoabsorption cross sections extracted from the ($p,p'$) data by means of the virtual photon method \cite{tam11}. 
The resulting E1 GSF is shown in Fig.~\ref{fig:e1} in comparison with three widely used models and with a GSF value at the neutron separation threshold deduced from experimental systematics over a wide mass range~\cite{cap09}.

The simplest model to describe the E1 GSF is a standard Lorentzian (SLO) function 
\begin{equation}
	\label{eqn:slo}
	f_{SLO}(E) = \frac{\sigma_r \Gamma_r}{3(\pi\hbar c)^2}
	\frac{\Gamma_r E}{(E^2-E_r^2)^2+(\Gamma_r E)^2},
\end{equation}
where $\sigma_r$ is the peak cross section, $E_r$ the centroid energy and $\Gamma_r$ the width of the IVGDR.

A more sophisticated model is the enhanced generalized Lorentzian (EGLO) model
\begin{eqnarray}
	\label{eqn:eglo}
	f_{EGLO}(E) &=& \frac{\sigma_r \Gamma_r}{3(\pi\hbar c)^2} \\ \nonumber
	& \times & \left[\frac{\Gamma_K(E,T)E}{(E^2-E_r^2)^2+(\Gamma_K(E,T)E)^2}\right. \\ \nonumber
	&+& \left.0.7 \, \frac{\Gamma_K(E=0,T)}{E_r^3}\right].
\end{eqnarray}
The EGLO consists of two terms~\cite{kop93}, a Lorentzian with an energy- and temperature-dependent width $\Gamma_K(E,T)$ and a term describing the shape of the low-energy part of the GSF. 
The temperature dependence is estimated within Fermi liquid theory \cite{pin66} 
\begin{equation}
	\label{eqn:egloWidth}
	\Gamma_K(E,T) = \chi(E)\frac{\Gamma_r}{E_r^2}(E^2+(2\pi T)^2) 	,
\end{equation}
where
\begin{equation}
	\label{eqn:egloChi}
	\chi(E) = \kappa + (1-\kappa) \frac{E-E_0}{E_r-E_0}
\end{equation}
is an empirical function with parameters $\kappa$ and $E_0$, where $\kappa$ is adjusted to reproduce the experimental E1 strength at a reference energy $E_0$~\cite{cap09}.

The SLO and EGLO models are both parametrizations of experimental data. 
In contrast, the modified Lorentzian model (MLO) is based on general relations between the GSF and the imaginary part of the nuclear response function
\begin{equation}
f_{MLO}(E) = \frac{\sigma_r \Gamma_r}{3(\pi\hbar c)^2}
 			  \frac{\Lambda(E,T)\Gamma(E,T)E}{(E^2-E_r^2)^2+(\Gamma(E,T)E)^2} ,
\end{equation}
where
\begin{equation}
	\Lambda(E,T) = \frac{1}{1-\exp(-E/T)}.
\end{equation}
The function $\Lambda(E,T)$ accounts for the enhancement of the GSF with increasing temperature. 
The width $\Gamma(E,T)$ within the MLO is calculated with microcanonically distributed initial states~\cite{plu99}.

The resulting predictions are shown in Fig.~\ref{fig:e1} as green (SLO), magenta (EGLO), and cyan (MLO) curves.
In the region around the maximum of the IVGDR all models provide a good description.
The high-energy tail of the IVGDR is well described by SLO and MLO while EGLO overestimates the photoabsorption cross sections.
The low-energy tail of the IVGDR exhibits strong fluctuations which complicate the comparison with smooth strength functions. 
For excitation energies down to about 8 MeV, MLO describes the average behavior fairly well while SLO(EGLO) are roughly consistent with the upper(lower) limits of the fluctuations but over(under)estimate the average cross sections.
Between 6 and 8 MeV a resonance-like structure dominates the GSF identified as the PDR in $^{208}$Pb \cite{pol12}.
This low-energy resonance is not included in the models.
Finally, the GSF value expected at neutron threshold ($S_n = 7.37$ MeV in the present case) from experimental systematics of neutron capture cross sections (black circle) is almost an order of magnitude smaller than the experimental strengths in the PDR.
However, this may be an artefact of the unusually low level density in the doubly magic nucleus $^{208}$Pb with corresponding strong fluctuations of individual strengths at energies close to the neutron threshold (note that the GSF values correspond to energy bins rather than to individual transitions for excitation energies above 7 MeV (cf.\ Tab. I in Ref.~\cite{pol12}). 

\subsection{M1 contribution}
\label{subsec:m1}

In addition to the $B$(E1) strengths measured in the $(p,pßprime)$ experiment, M1 cross sections at $\Theta = 0^\circ$ are provided in Tab.\ I of Ref.~\cite{pol12}.
These are concentrated between 7 and 9~MeV and represent the spinflip M1 resonance \cite{hey10}.
We note that  an additional M1 transition to a $1^+$ state at 5.844 MeV is known (see Ref.~\cite{sch10} and references therein) but omitted here because it is of dominant isoscalar nature \cite{mue85}. 
Recently a method utilizing isospin symmetry has been presented to relate the spinflip M1 cross sections to those of Gamow-Teller excitations studied with the $(p,n)$ reaction and extract the spin-M1 matrix elements \cite{bir16}.
Assuming dominance of the spin-isovector part of the electromagnetic M1 operator reduced $B$(M1) transition strengths  can be extracted.
In the resonance region these agree well \cite{bir16} with studies using electromagnetic probes \cite{koe87,las88}.
At excitation energies above 8 MeV, where previous experiments had limited sensitivity \cite{koe87}, additional $B$(M1) strength is found in the $(p,p^\prime)$ data which raises the total strength of the spinflip M1 resonance by about 20\%.

\begin{figure}
	\centering
	\includegraphics[width=8.6cm]{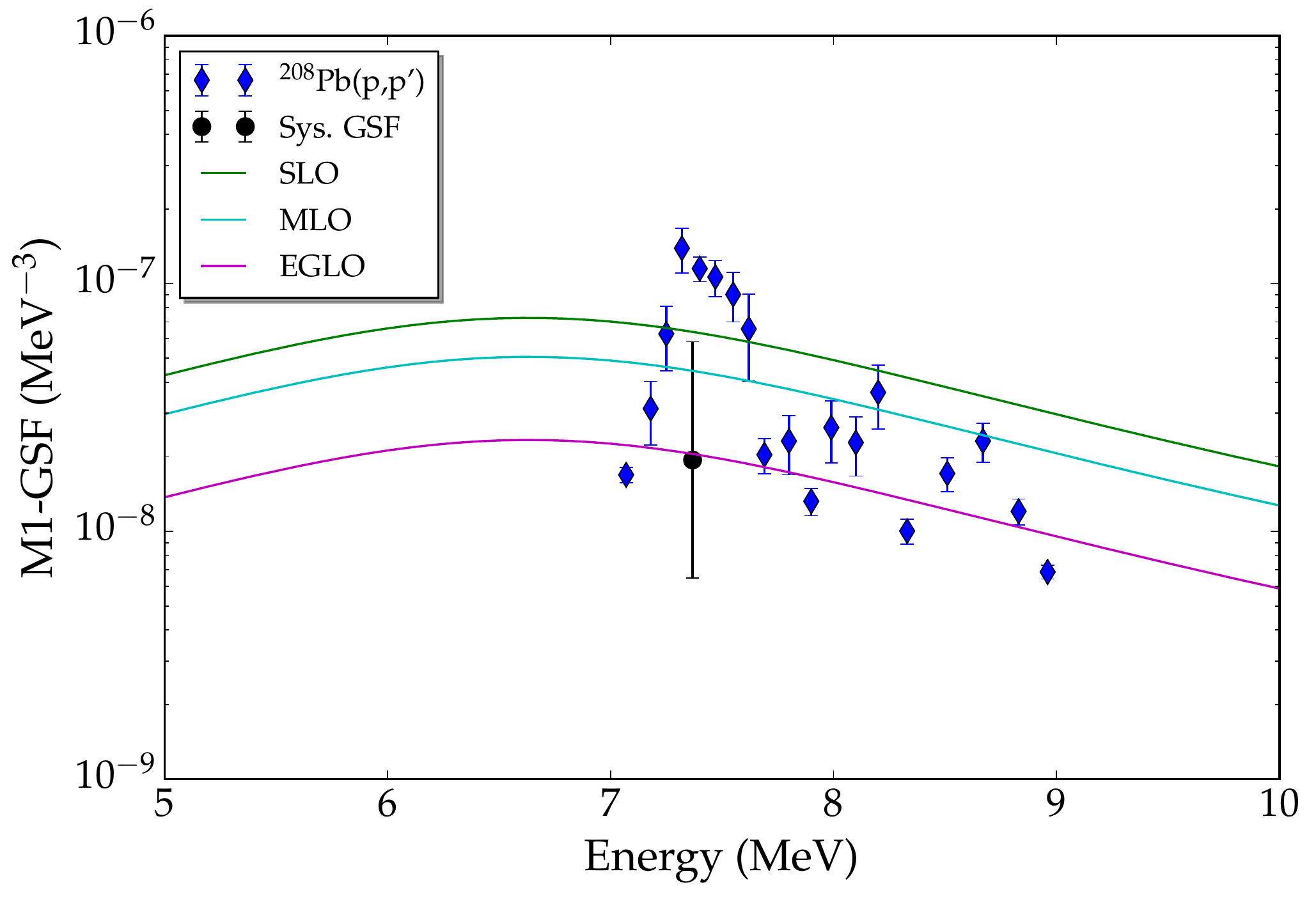}
	\caption{(Color online) Same as Fig.~\ref{fig:e1} but for the M1 component of the GSF.}
	\label{fig:m1}
\end{figure}
Figure \ref{fig:m1} displays the corresponding M1 GSF in comparison with SLO, EGLO, and MLO model predictions for $^{208}$Pb.
The M1 GSF model results are derived from the E1 models discussed above in the following way \cite{cap09}
\begin{equation}
	f^{M1}(E) =  \frac{f^{E1}(S_n)}{\mathcal{R}}
	\frac{\Phi^{M1}(E)}{\Phi^{M1}(S_n)} 
	\end{equation}
and
\begin{equation}
	\mathcal{R} = \frac{f^{E1}(S_n)}{f^{M1}(S_n)} = 0.0588 A^{0.878}, 
\label{eq:r}
\end{equation}
where $\Phi^{M1}(E)$  is a SLO parametrization of the spinflip M1 resonance with energy centroid $E_r = 41 \cdot A^{-1/3}$ and $\Gamma_r = 4$ MeV \cite{kop87}.
The mass dependence of the ratio $\mathcal{R}$ in Eq.~(\ref{eq:r}) is valid for nuclei with $S_n \approx 7$~MeV.
Thus, it should be a good approximation for $^{208}$Pb.

The comparison in Fig.~\ref{fig:m1} indicates that the theoretical GSF values near maximum are of magnitudes roughly comparable to the data.
However, the assumed resonance properties represent a poor approximation of the data.
The theoretical maxima are about 500 keV too low and the experimental width is grossly overestimated.
As a result, the predicted total strengths of the spinflip M1 resonance exceed the experimental value $\sum B({\rm M1}) = 20.5(13) \, \mu_N^2$ \cite{bir16} by factors ranging from two (EGLO) to five (SLO).

\subsection{E2 contribution}
\label{subsec:e2}

\begin{figure}
	\centering
	\includegraphics[width=8.6cm]{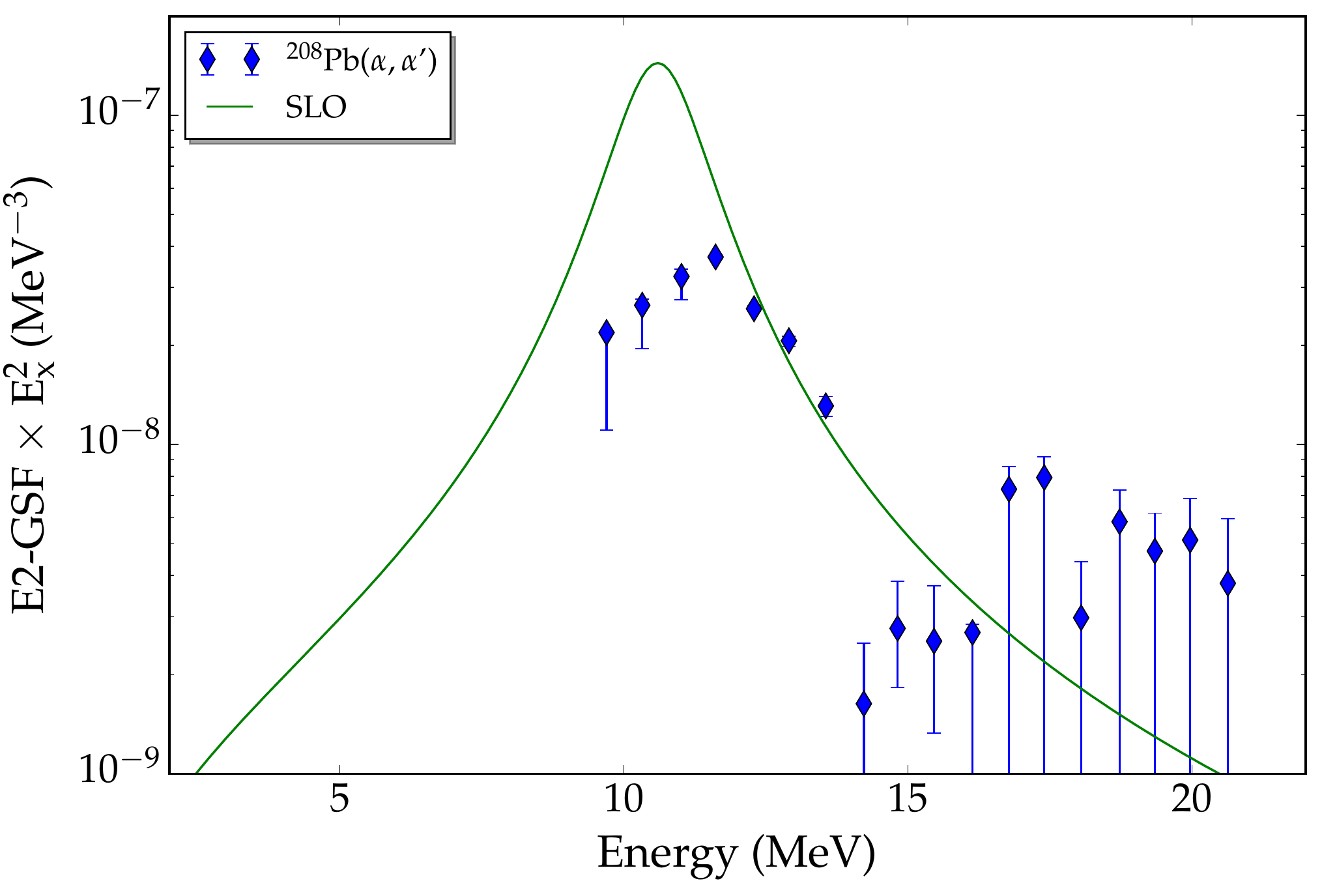}
	\caption{(Color online) E2 GSF deduced from Ref.~\cite{you04} in comparison to the SLO model with parameters from Ref.~\cite{pre84}.
The GSF was multiplied by $E_x^2$ to make the units comparable to the E1 and M1 GSF.}
	\label{fig:e2}
\end{figure}

The E2 contribution to the GSF was estimated using $(\alpha$,$\alpha ')$ data obtained at the Texas A\&M K500 superconducting cyclotron, College Station, Texas, USA~\cite{you04}. 
In this experiment several isotopes including $^{208}$Pb were investigated using alpha particles with an energy of 240~MeV. 
The data was taken in an excitation energy region of 10 to 55~MeV where isoscalar E0, E1, E2 and E3 strength distributions were extracted with the aid of a MDA. 
The resulting E2 strength distribution exhausted $100\pm15\%$ of the energy weighted sum rule (EWSR). 
Using this data the B(E2) strength distribution was obtained and converted to the E2 GSF shown in Fig.~\ref{fig:e2}.
The solid line shows a global parametrization of the E2 giant resonance \cite{pre84} suggested in earlier RIPL versions.

\subsection{Total GSF and comparison with Oslo data}
\label{sec:comparison}

\begin{figure}
	\centering
	\includegraphics[width=8.6cm]{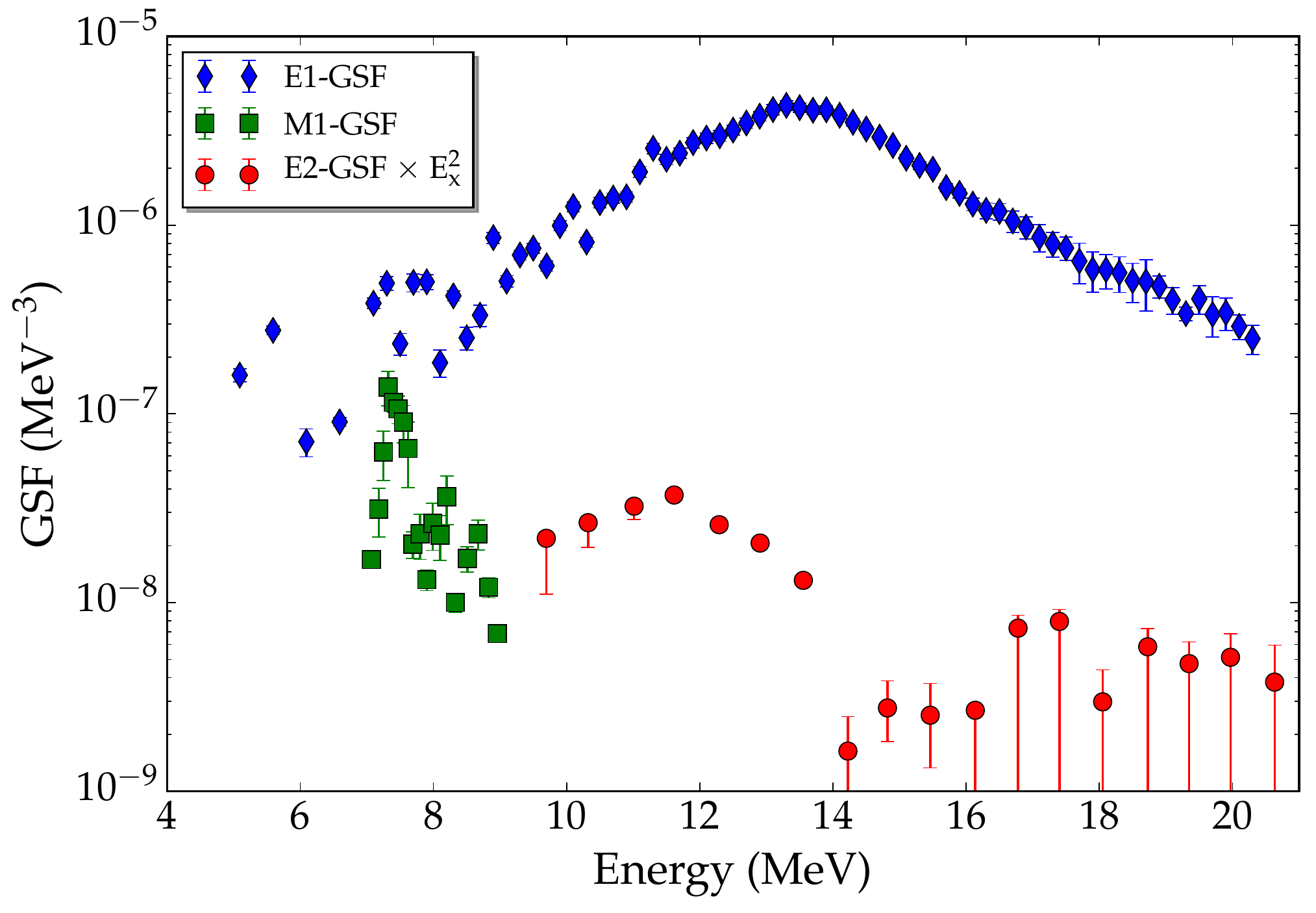}
	\caption{(Color online) Comparison of E1, M1 and E2 contributions to the GSF of $^{208}$Pb.}
	\label{fig:gsf}
\end{figure}

Figure~\ref{fig:gsf} summarizes  the E1, M1 and E2 contributions to the total GSF.
 As can be seen, the dominant contribution stems from E1 transitions. 
The M1 contribution is of the order of a few percent for excitation energies above 8 MeV and reaches at most 10-30\% in the peak of the resonance around $S_n$. 
The E2 contribution is of comparable magnitude to M1 but located at higher excitation energies. 
Because of the simultaneous strong rise of the E1 part in the IVGDR energy region the E2 contribution to the GSF at the maximum of the E2 resonance is about 1\% only.  

\begin{figure}
	\centering
	\includegraphics[width=8.6cm]{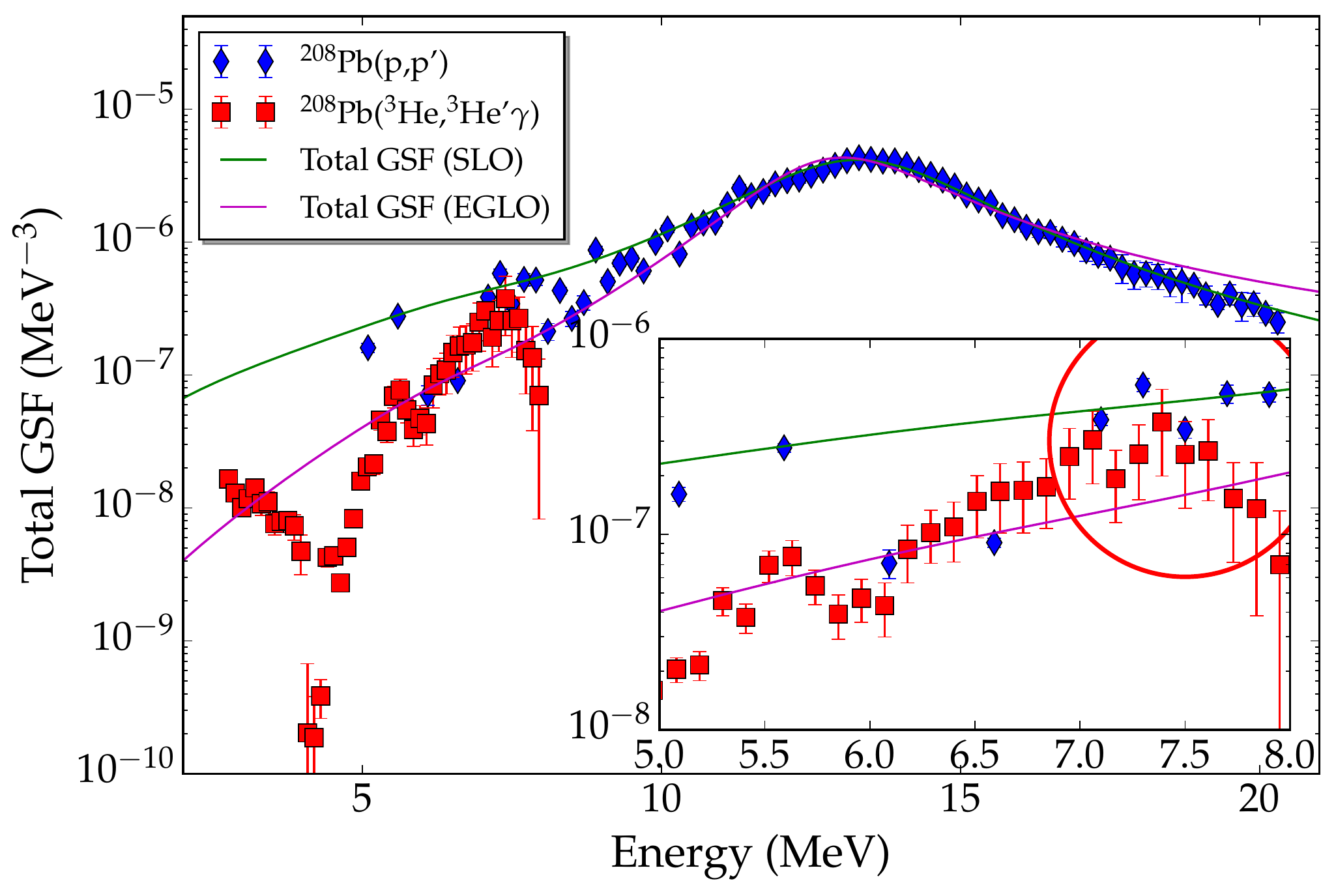}
	\caption{(Color online) 
	Total GSF of $^{208}$Pb from the ($p,p'$) data \cite{tam11,pol12} in comparison to the reanalyzed \cite{gutpc} results fromm the Oslo experiment \cite{sye09}.
The inlet shows and expanded view of the low-energy region $5-8$ MeV.	
}
	\label{fig:totGSF}
\end{figure}

The total GSF summing all contributions is displayed in Fig.~\ref{fig:totGSF} (blue diamonds)  and compared with data derived with the Oslo method from a $^{208}$Pb($^3$He,$^3$He$^\prime \gamma$) experiment (red circles) \cite{sye09}.
The data set has been reanalyzed recently~\cite{gutpc}. 
The main changes are new, updated response functions for the CACTUS detector array and an improved error estimate in the simultaneous extraction of level density and $\gamma$ strength from the primary $\gamma$-ray spectra. 
The initial excitation energy range used for the reanalysis was $4.75 \leq E_{i} \leq 7.95$ MeV, and the applied low-$E_{\gamma}$ threshold was 2.65 MeV. 
For consistency with the previous work, the level density has been normalized to the $p$-wave resonance data of RIPL-2, see Tab. I in Ref.~\cite{sye09}. 
Further, the $\gamma$ strength has been normalized to recent ($\gamma$,n) data by Kondo \textit{et al.}~\cite{kon12} and also compared to older data \cite{har64,vey70}. 
This is considered as the low-limit estimate of the $\gamma$ strength from the $^{3}$He-induced reaction.

There are overlapping results from both experiments in the energy region between 5 and 8 MeV (see inlet of Fig.~\ref{fig:totGSF}).
The GSF derived from  the $(p,p^\prime)$ data is systematically higher in the PDR region although they seem still compatible within error bars in the peak region around the neutron threshold.
Between 6 and 7 MeV consistent results are found while below 6 MeV the strong transitions observed in Ref.~\cite{pol12} exceed the average $\gamma$ strength in the Oslo data by factors 4 to 5.
However, one should be aware that single transitions are analyzed for excitation energies $E_{\rm x} < 7$ MeV \cite{pol12} and the level density of $1^-$ states excited from the ground state is probably too low to discuss an average behavior in the PDR region.
Rather the upward GSF is dominated by Porter-Thomas intensity fluctuations.

\section{Level density of $^{208}$Pb}
\label{sec:nld}

Level density of $1^-$ states in the excitation energy region from 9 to 12.5~MeV was determined from ($p,p^\prime$) data \cite{pol14} using a fluctuation analysis~\cite{han90}. 
However, the LDs for $^{208}$Pb derived from the Oslo method represent a different spin window depending on the specific reaction.
Thus all results are converted to total level densities using Fermi gas models \cite{sch00}. 
This can be achieved using the following equation
\begin{equation}
	\rho^{tot}(E) = \frac{2\rho(E,J,\Pi)}{f(J)},
\end{equation}
where $\rho^{tot}(E)$ is the total level density at energy $E$ and $\rho(E,J,\Pi)$ is the level density for transitions with spin $J$ and parity $\Pi$. 
The function $f(J)$ is the so-called spin distribution function defined as
\begin{equation}
	f(J) = \frac{2J+1}{2\sigma^2}\exp
	\left[-\frac{(J+\frac{1}{2})^2}{2\sigma^2}\right],
\end{equation}
where $\sigma$ is the spin cutoff parameter. 
Since the spin cutoff depends on the parameters of the Fermi gas model one has to investigate the model dependence.
For this purpose, we considered three parameter sets derived within the backshifted Fermi gas model (BSFG)  approach \cite{dil73}.
These include the one used in the original analysis of the Oslo experiment \cite{sye09}, a global set recommended in RIPL-3 \cite{cap09}, and the parametrization of Ref.~\cite{rau97} developed for $s$-process reaction network calculations, which  has been shown to provide a good description of LD for many nuclei near the valley of stability \cite{pol14,kal06,kal07,has08}.
 
\begin{figure}
	\centering
	\includegraphics[width=8.6cm]{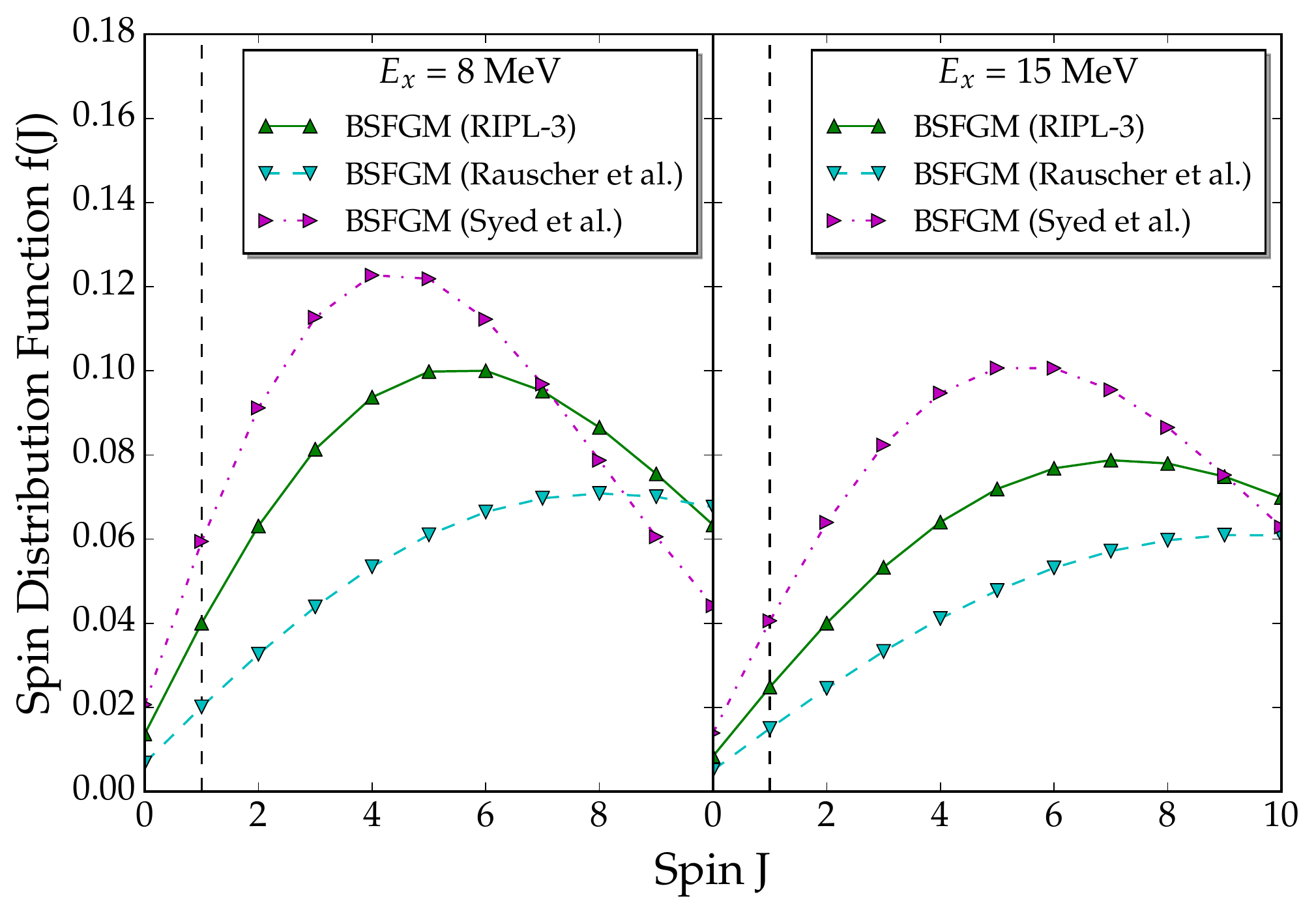}
	\caption{(Color online) Spin distribution functions of LDs in $^{208}$Pb at excitation energies of 8 and 15~MeV from BSFG model predictions with the parameters of Ref.~\cite{sye09} (magenta dashed-dotted lines), Ref.~\cite{cap09} (green solid lines), and Ref.~\cite{rau97} (cyan dashed lines).
The vertical dashed lines indicate the values for spin $J = 1$.}
	\label{fig:spinDist}
\end{figure}

Figure~\ref{fig:spinDist} shows the spin distribution functions from the three different parametrizations at excitation energies of 8 and 15 MeV, which show significant differences.
The values  for $J = 1$ are indicated by the vertical dashed lines.

\begin{figure}[b]
	\centering
	\includegraphics[width=8.6cm]{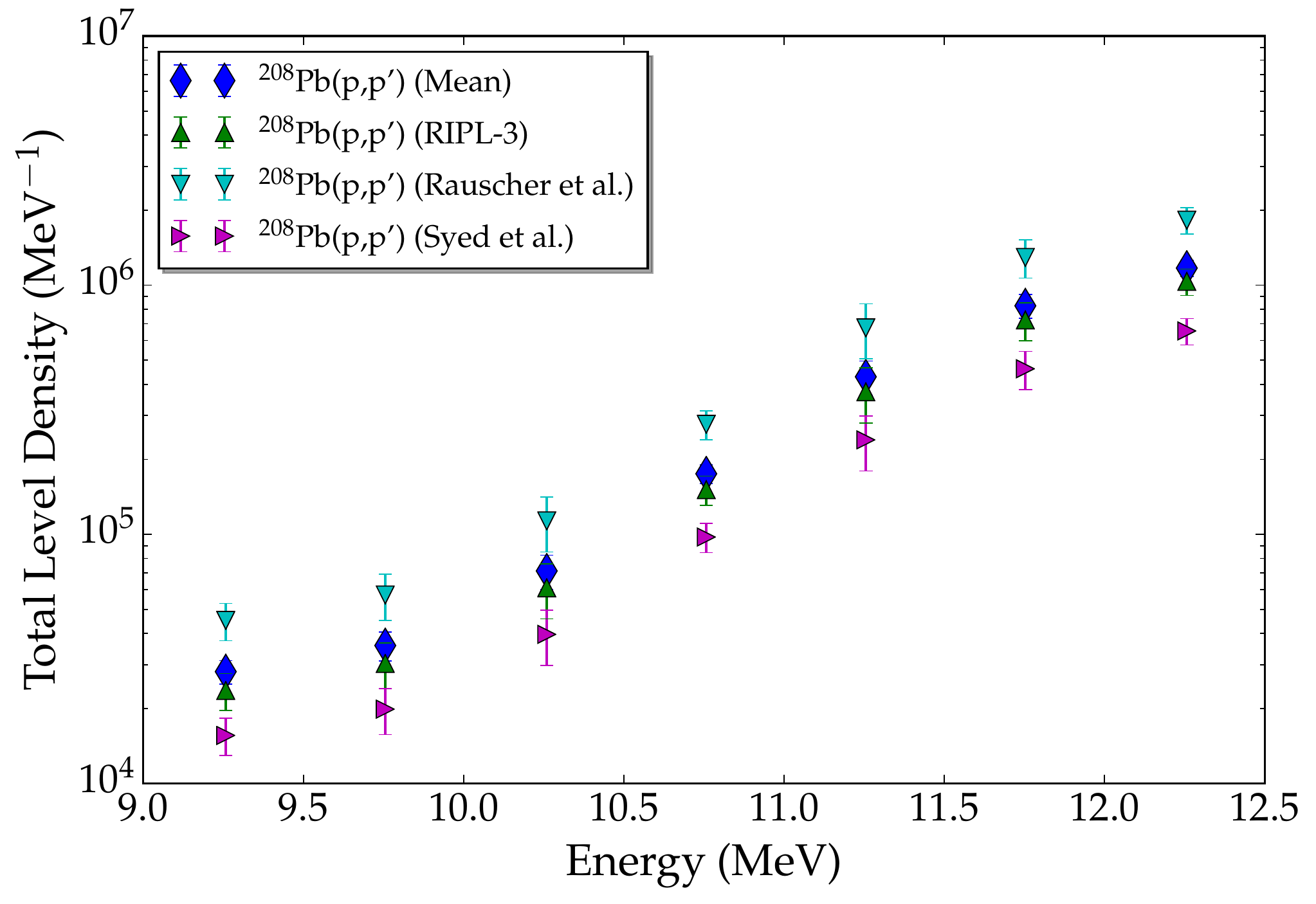}
	\caption{(Color online) 
Total level density of $^{208}$Pb between 9 and 12.5 MeV obtained with the spin distribution functions of Ref.~\cite{sye09} (rightpointing magenta triangles), Ref.~\cite{cap09} (green upward triangles), and Ref.~\cite{rau97} (cyan downward triangles).
Mean values averaged over the three models are shown as blue diamonds.}
	\label{fig:BSFG}
\end{figure}

Figure \ref{fig:BSFG} presents the resulting total LDs in $^{208}$Pb from the three models for an excitation energy range $E_{\rm x} = 9 -12.5$ MeV.
The absolute values depend on the chosen parametrization (cf.\ Fig.~\ref{fig:spinDist}) but their ratio shows limited variation over the studied energy region, i.e., all three models predict a similar energy dependence.
Therefore, we use averaged values for the comparison with the Oslo data (blue diamonds).
The error bars include the model dependence which actually dominates over the uncertainties in the extraction of the $1^-$ LD discussed in Ref.~\cite{pol14}.   

\begin{figure}
	\centering
	\includegraphics[width=8.6cm]{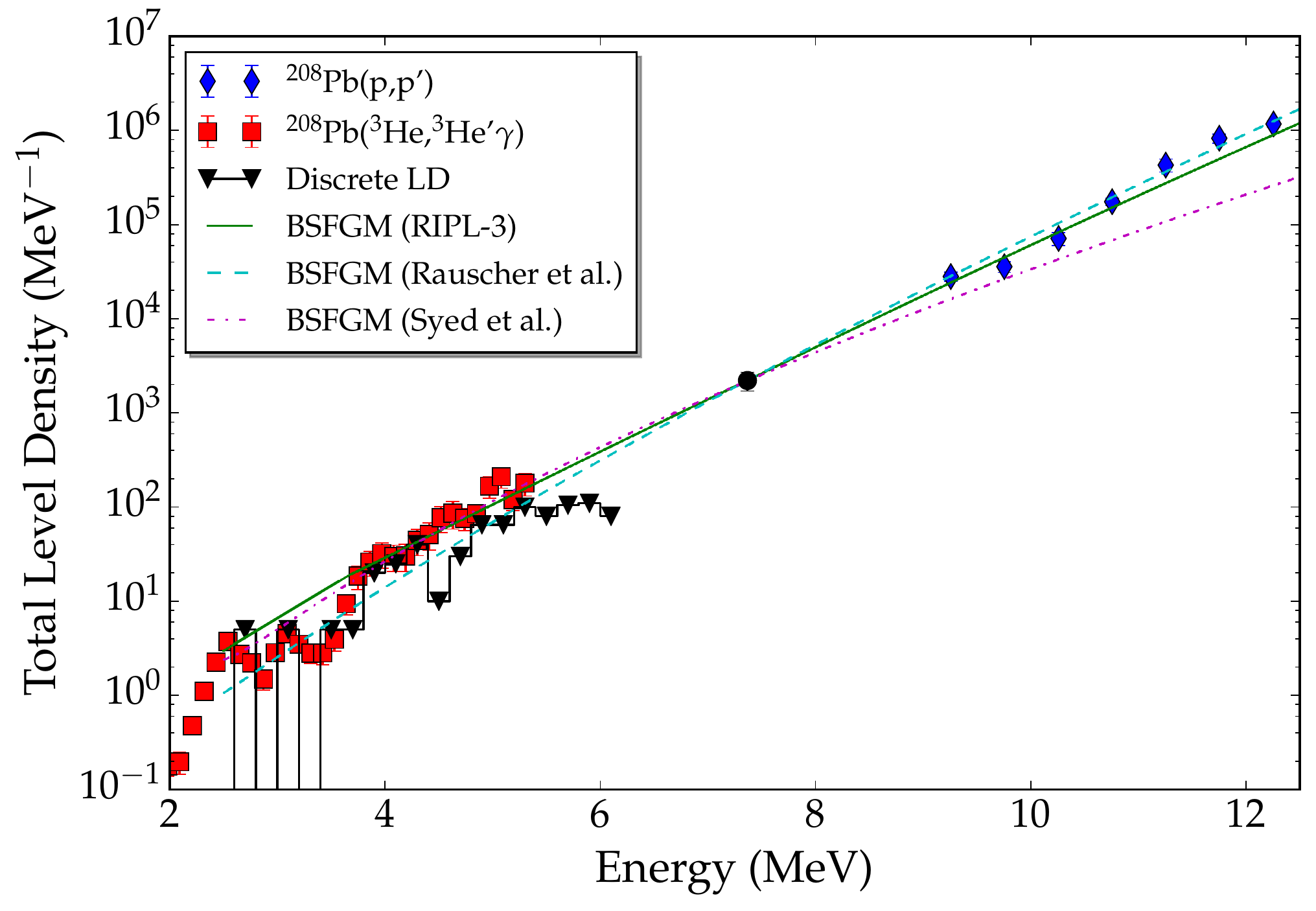}
	\caption{(Color online) 
Total LD from the ($p,p'$) data \cite{tam11,pol12} in comparison to the reanalyzed \cite{gutpc} results from the Oslo experiment \cite{sye09}.
The black downward triangles are results from from counting the levels identified in Ref.~\cite{heu16} in 200 keV bins.
The magenta dashed-dotted, green solid, and cyan dashed lines are  BSFG model predictions with the parameters of Ref.~\cite{sye09}, Ref.~\cite{cap09}, and Ref.~\cite{rau97}, respectively. 
}
	\label{fig:NLD}
\end{figure}

The comparison with the Oslo results (red squares) is finally presented in Fig.~\ref{fig:NLD}.
The value at neutron threshold (black circle) is deduced from $p$-wave resonance neutron capture converted to a total LD with the aid of the RIPL-3 BSFG parametrization.
The black downward triangles denote the results from level counting from a recent study claiming essentially complete spectrocopy up to 6.2 MeV \cite{heu16}.
Indeed, the LD agrees well with the Oslo result up to about 5 MeV.
For $E_{\rm x} > 5$ MeV, the LDs deduced from Ref.~\cite{heu16} are approximately constant indicating that an increasing amount of levels is missing (approximately a factor of two at 6 MeV).

The magenta dashed-dotted, green solid, and cyan dashed lines are the BSFG model predictions with the parameters of Ref.~\cite{sye09}, Ref.~\cite{cap09}, and Ref.~\cite{rau97}, respectively. 
The models are normalized to the data point at $S_n$ with factors 1.15, 2.18, and 0.52. 
However, absolute values for the RIPL-3 parametrization are obtained by normalizing to $s$-wave neutron capture resonance spacings. 
As pointed out in Ref.~\cite{sye09}, the data are rather poor in $^{208}$Pb and one should rather normalize to the $p$ wave spacings, i.e., the solid curve is absolute.  
The energy dependence of the BFSG model shows differences over the wide energy range spanned by the two data sets.
Remarkably, the RIPL-3 parameter set, whose predicions of the total LD are closest to the mean value (cf.~Fig.~\ref{fig:BSFG}), provides a consistent description of all data.  

\section{Summary and conclusions}
\label{sec:summary}

The main aim of this work was to determine the E1, M1 and total GSF of $^{208}$Pb for tests of models recommended in the RIPL-3 data base as well as to study the BA hypothesis by comparison with decay data obtained with the Oslo method. 
It is shown that the E1 GSF can be described well by the SLO and MLO models in the GDR region. 
In the low-energy region strong fluctuations occur, so that no particular model can be favored.  
The average behavior of the low-energy tail of the IVGDR is probably best described by the MLO model.
However, none of models includes the PDR and thus the predictive power at low excitation energies is generally limited.

The presently recommended parametrization of the spinflip M1 resonance provides only a poor description of the $^{208}$Pb data.
Although the absolute magnitude of the resonance maximum is reproduced within a factor of 2 to 3, the width of the M1 GSF is strongly overestimated.
As a result the $B$(M1) strengths is predicted by the empirical models are too large by factors 2 to 5. 
Since the excitation energy ranges of the spinflip M1 resonance and the PDR overlap in heavy nnuclei, this has a strong impact on attempts to extract model parameters for the PDR contribution in decay experiments.
Clearly, more data are needed to establish the systematics of the poorly known spinflip M1 resonance in heavy nuclei. 
The method presented in Ref.~\cite{bir16} promises experimental information from the $(p,p^\prime)$ data on spherical \cite{kru15,iwa12} as well as deformed \cite{vnc16,kru13} nuclei.

The comparison of the present GSF derived from ground-state absorption with the Oslo results shows larger values in the PDR energy region, where both data sets overlap.
However, the fluctuations of the GSF are very strong due to the anomalously small level densities in the closed-shell nucleus $^{208}$Pb, which prevents conclusions on a possible violation of the BA hypothesis in the PDR energy region.
Here, tests in open-shell nuclei with higher level densities are required and a corresponding study \cite{vnc16} is underway.  

Total level densities for $^{208}$Pb  were derived from fluctuations of the high-resolution $(p,p^\prime)$ cross sections in the IVGDR energy region and compared to those from the Oslo method covering much lower energies.
Using the BSFG model parameters suggested by RIPL-3 to convert the experimental partial-spin results to total level densities and to describe their energy dependence, remarkable agreement between the two results is obtained.
This provides an independent confirmation of the approach \cite{sch00} to separate GSF and LD in Oslo-type data.      
  
\begin{acknowledgments}

Discussions with M.~Guttormsen, A.~C.~Larsen, A.~Richter, and S.~Siem are gratefully acknowledged.  
We are indebted to M.~Guttormsen and A.~C.~Larsen for providing us with the revised analysis of the data of Ref.~\cite{sye09}.
This work has been supported by the DFG under contract SFB 1245 and by the JSPS, KAKENHI Grant Number JP14740154.

\end{acknowledgments}


\end{document}